\documentclass[prb,showpacs,floatfix,twocolumn]{revtex4}
\usepackage{amssymb}
\usepackage{txfonts}
\usepackage{graphicx}
\usepackage{dcolumn}
\usepackage{float}

\begin{document}



\title{Electronic properties of 3\emph{d} transitional metal pnictides : A comparative study by optical spectroscopy}
\author{B. Cheng}
\author{B. F. Hu}
\author{R. Y. Chen}
\author{G. Xu}
\author{P. Zheng}
\author{J. L. Luo}
\author{N. L. Wang}

\affiliation{ Institute of Physics, Chinese Academy of Sciences,
Beijing 100080, People's Republic of China}
%


\begin{abstract}
Single-crystalline KFe$_{2}$As$_{2}$ and CaT$_{2}$As$_{2}$ (T = Fe,
Co, Ni, Cu) are synthesized and investigated by resistivity,
susceptibility and optical spectroscopy. It is found that
CaCu$_{2}$As$_{2}$ exhibits a similar transition to the lattice
abrupt collapse transitions discovered in
CaFe$_{2}$(As$_{1-x}$P$_{x}$)$_{2}$ and
Ca$_{1-x}$\emph{Re}$_{x}$Fe$_{2}$As$_{2}$ (\emph{Re} = rare-earth
element). The resistivity of KFe$_{2}$As$_{2}$ and CaT$_{2}$As$_{2}$
(T = Fe, Co, Ni, Cu) approximately follows the similar \emph{T
$^{2}$} dependence at low temperature, but the magnetic behaviors
vary with different samples. Optical measurement reveals the optical
response of CaCu$_{2}$As$_{2}$ is not sensitive to the transition at
50 K, with no indication of development of a new energy gap below
the transition temperature. Using Drude-Lorentz model, We find that
two Drude terms, a coherent one and an incoherent one, can fit the
low-energy optical conductivity of KFe$_{2}$As$_{2}$ and
CaT$_{2}$As$_{2}$ (T = Fe, Co, Ni) very well. However, in
CaCu$_{2}$As$_{2}$, which is a \emph{sp}-band metal, the low-energy
optical conductivity can be well described by a coherent Drude term.
Lack of the incoherent Drude term in CaCu$_{2}$As$_{2}$ may be
attributed to the weaker electronic correlation than
KFe$_{2}$As$_{2}$ and CaT$_{2}$As$_{2}$ (T = Fe, Co, Ni). Spectral
weight analysis of these samples indicates that the unconventional
spectral weight transfer, which is related to Hund's coupling energy
\emph{J$_{H}$}, is only observed in iron pnictides, supporting the
viewpoint that \emph{J$_{H}$} may be a key clue to seek the
mechanism of magnetism and superconductivity in pnictides.
\end{abstract}

\pacs{75.30.Cr, 74.70.Xa, 74.25.nd}

\maketitle

\section{\label{sec:level2}Introduction}

The discovery of iron-based superconductors in 2008 has triggered a
new wave of researches on the veiled mechanism of
high-\emph{T$_{c}$} superconductivity and its interplay with
magnetism.\cite{Fe1} Similar to the cuprates, the parent compounds
of many iron-based superconductors exhibit long-range
antiferromagnetism at low temperature, and doping electrons or holes
will suppress magnetism and introduce
superconductivity.\cite{Fe-M,Fe2,Fe3} In the 1111 family,
\emph{T$_{c}$} reaches up to 55 K,\cite{Fe4} which is much higher
than the value predicted by traditional electron-phonon coupling
theory. It is widely believed that the superconductivity in iron
pnictides has an unconventional origin.

Besides the work conducted at raising \emph{T$_{c}$} in iron-based
superconductors, efforts have also been made for the exploration of
some new compounds which have similar crystal structures with
iron-based superconductors, such as a complete substitution of Fe by
Cr, Mn, Co, Ni or Cu in iron
pnictides.\cite{Mn-1,Mn-2,Cr,Co5,Fe-Co,Cu} These complete
substitutions significantly affect band structures and modify the
topological shapes of Fermi surfaces compared to iron pnictides, and
may introduce new physical phenomena like some complete
substitutions in the cuprates. BaCo$_{2}$As$_{2}$ exhibits a
paramagnetic behavior above 1.8 K, and the enhancement of
susceptibility relative to free-electron systems indicates that it
is close to a magnetic quantum critical point.\cite{Co5}
BaNi$_{2}$As$_{2}$ displays a first-order crystal structure phase
transition at 130 K. Optical and band structure calculation
investigation reveal that its several small Fermi surface sheets
contributed dominantly from the As-As bonding and Ni-As antibonding
are removed across the transition, resulting in a new energy gap
around 5000 cm$^{-1}$.\cite{Fe-Co,Ni} However, in contrast with the
positive magnetic susceptibility in BaCo$_{2}$As$_{2}$ and
BaNi$_{2}$As$_{2}$, SrCu$_{2}$As$_{2}$ exhibits diamagnetism in the
measurable temperature range.\cite{Cu} Local-density approximation
calculations for BaCu$_{2}$As$_{2}$ and SrCu$_{2}$As$_{2}$
substantiate that they are \emph{sp}-band metals. The 3\emph{d}
electrons of Cu are mainly located at 3 eV and higher binding energy
and are therefore chemically inert with little contribution to the
states near the Fermi energy.\cite{Cu-DOS}

On the basis of these interesting researches, we have synthesized a
series of single-crystalline samples, including KFe$_{2}$As$_{2}$,
CaFe$_{2}$As$_{2}$, CaCo$_{2}$As$_{2}$, CaNi$_{2}$As$_{2}$ and
CaCu$_{2}$As$_{2}$. From KFe$_{2}$As$_{2}$ to CaNi$_{2}$As$_{2}$,
according to the balance of chemical valances, the nominal number of
3\emph{d} electrons on per transition metal ion varies from 5.5 to
8, and these 3\emph{d} electrons are considered as the main
contribution to the density of states near the Fermi level and will
participate in various low-energy physical processes. However,
CaCu$_{2}$As$_{2}$ is a \emph{sp}-band metal as BaCu$_{2}$As$_{2}$
and SrCu$_{2}$As$_{2}$. All ten 3\emph{d} electrons of Cu ion are
located below the Fermi level with little contribution to the
density of states at Fermi energy. The different configuration of
\emph{3d} electrons between CaCu$_{2}$As$_{2}$ and CaT$_{2}$As$_{2}$
(T = Fe, Co, Ni) provides a good platform to investigate the
difference of low-energy optical response with different types of
electron.

In this article, we report our investigation on these samples by
resistivity, susceptibility and optical spectroscopy. We find that
CaCu$_{2}$As$_{2}$ undergoes a transition at 50 K, which is similar
to the lattice abrupt collapse transitions discovered in
CaFe$_{2}$(As$_{1-x}$P$_{x}$)$_{2}$ and
Ca$_{1-x}$\emph{Re}$_{x}$Fe$_{2}$As$_{2}$ (\emph{Re} = rare-earth
element).\cite{Ca-P,Ca-Re} However, Optical measurements reveal that
the optical response of CaCu$_{2}$As$_{2}$ is not very sensitive to
the transition at 50 K. The temperature dependent resistivity of
KFe$_{2}$As$_{2}$ and CaT$_{2}$As$_{2}$ (T = Fe, Co, Ni, Cu) shows
some common features. Above 200 K, all resistivity exhibit
linear-\emph{T} dependence and below 50 K approximately follow
\emph{T}$^{2}$ dependence. We use Drude-Lorentz model to analyze
$\sigma$$_{1}$($\omega$) of KFe$_{2}$As$_{2}$ and CaT$_{2}$As$_{2}$
(T = Fe, Co, Ni, Cu). We find that the low-energy
$\sigma$$_{1}$($\omega$) of KFe$_{2}$As$_{2}$ and CaT$_{2}$As$_{2}$
(T = Fe, Co, Ni) can be well reproduced by two Drude terms. However,
in CaCu$_{2}$As$_{2}$, only one coherent Drude term can well account
for the low-energy $\sigma$$_{1}$($\omega$). Lack of the incoherent
Drude term in CaCu$_{2}$As$_{2}$ may be attributed to the weaker
electronic correlation compared to KFe$_{2}$As$_{2}$ and
CaT$_{2}$As$_{2}$ (T = Fe, Co, Ni). We also perform the spectral
weight analysis on these samples. We find the unconventional
spectral weight transfer which is related to Hund's coupling energy
\emph{J$_{H}$} is only observed in iron pnictides, indicating that
\emph{J$_{H}$} may play an important role in the mechanism of
magnetism and superconductivity in pnictides.

\section{\label{sec:level2}EXPERIMENTAL DETAIL}

High-quality KFe$_{2}$As$_{2}$, CaFe$_{2}$As$_{2}$,
CaCo$_{2}$As$_{2}$, CaNi$_{2}$As$_{2}$ and CaCu$_{2}$As$_{2}$ single
crystals are grown by self-flux method. Resistivity measurements are
performed on Quantum Design physical property measurement system
(PPMS). Dc susceptibilities are measured as a function of
temperature using Quantum Design instrument superconducting quantum
inference device (SQUIT-VSM). The optical reflectance measurements
with E $\parallel$ \emph{ab} plane are performed on Bruker IFS 80v
and 113v spectrometers in the frequency range from 30 to 25000
cm$^{-1}$. In situ gold and aluminum overcoating technique are used
to obtain the reflectivity \emph{R($\omega$)}. The real part of
conductivity $\sigma$$_{1}$($\omega$) is obtained by the
Kramers-Kronig transformation of \emph{R($\omega$)}. The density of
states of CaCu$_{2}$As$_{2}$ is calculated by density functional
theory implemented in the VASP code .

\section{\label{sec:level2} RESULTS AND DISSCUSION}

\begin{figure}[b]
\scalebox{0.5} {\includegraphics [bb=600 25 -3cm 11cm]{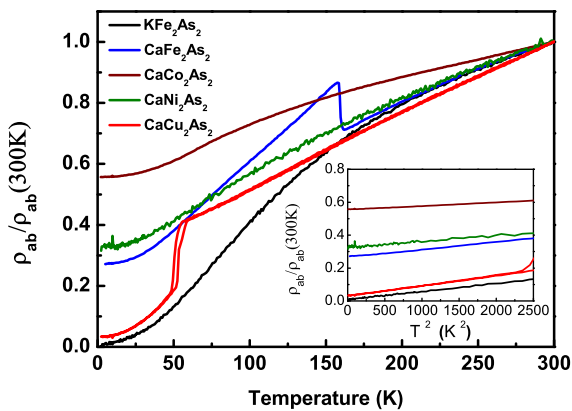}}
\caption{(color online)Temperature dependent resistivity of
KFe$_{2}$As$_{2}$ and CaT$_{2}$As$_{2}$ (T=Fe, Co, Ni, Cu) in zero
field for \emph{\textbf{I} $\parallel$ ab}. The inset plots
resistivity as a function of \emph{T $^{2}$}}
\end{figure}

Figure 1 shows the temperature-dependent resistivity of
KFe$_{2}$As$_{2}$ and CaT$_{2}$As$_{2}$ (T = Fe, Co, Ni, Cu) with
\emph{\textbf{I} $\parallel$ ab}. For the convenience of comparison,
we have normalized $\rho$\emph{$_{ab}$}(\emph{T}) to
$\rho$\emph{$_{ab}$}(300 K). All five samples show a metallic
behavior in the measured temperature range. The resistivity of
CaFe$_{2}$As$_{2}$ and KFe$_{2}$As$_{2}$ is basically consistent
with some earlier reports about CaFe$_{2}$As$_{2}$ and
KFe$_{2}$As$_{2}$.\cite{Ca1,K} In contrast with BaNi$_{2}$As$_{2}$,
the resistivity of CaNi$_{2}$As$_{2}$ decreases smoothly with
lowering temperature and does not exhibit any anomaly, indicating
that the similar structure phase transition which occurs in
BaNi$_{2}$As$_{2}$ is absent in CaNi$_{2}$As$_{2}$.\cite{Ni}
Furthermore, earlier transport study of SrCu$_{2}$As$_{2}$ shows
that it does not undergo any phase transition.\cite{Cu} However,
shown in figure 1, the resistivity of CaCu$_{2}$As$_{2}$ displays a
sharp drop at 50 K. With decreasing and then increasing temperature,
a notable hysteresis is observed, which provides evidence for the
occurrence of a first-order phase transition in CaCu$_{2}$As$_{2}$.

The temperature dependent resistivity of these five samples shares
some common features. Above 200 K, the resistivity of all five
sample basically obey the linear-temperature dependence. With
decreasing temperature, the resistivity of KFe$_{2}$As$_{2}$,
CaCo$_{2}$As$_{2}$ and CaNi$_{2}$As$_{2}$ begins to deviate from
\emph{T}-dependent behavior slowly. Further decreasing temperature,
the resistivity of all samples deviates from linear-temperature
dependent behavior completely and begins to show a
\emph{T}$^{2}$-dependence below 50 K. The inset of Figure 1 depicts
the resistivity of these five sample as a function of \emph{T}$^{2}$
below 50 K, and the linear relation is clearly seen.

\begin{figure}[t]
\scalebox{0.5} {\includegraphics [bb=330 30 6cm 11.5cm]{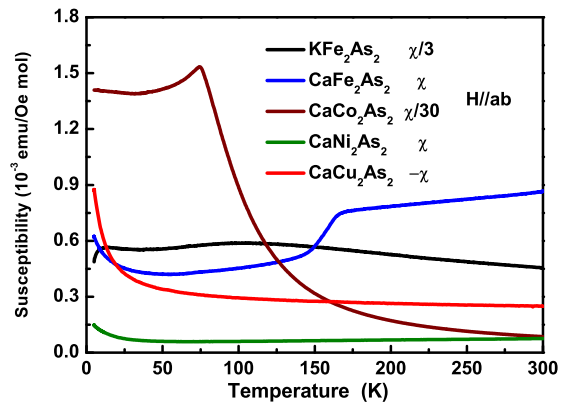}}
\caption{(Color online) Temperature dependent magnetic
susceptibilities of KFe$_{2}$As$_{2}$ and CaT$_{2}$As$_{2}$ (T=Fe,
Co, Ni, Cu) with \textbf{\emph{H}}$\parallel$ \emph{ab}.
Susceptibilities of these samples are measured at 1 T}
\end{figure}

The DC susceptibilities of KFe$_{2}$As$_{2}$ and CaT$_{2}$As$_{2}$
(T = Fe, Co, Ni, Cu) are shown in Fig. 2. $\chi$(T) has been
normalized to a proper integer for the convenience of comparison.
The susceptibility of CaFe$_{2}$As$_{2}$ reveals a sharp step-like
drop at 165 K, indicative of the spin density wave (SDW) transition.
Although KFe$_{2}$As$_{2}$ does not exhibit static long-range
magnetic ordering, its susceptibility is about three times larger
than CaFe$_{2}$As$_{2}$ and nearly shows no temperature
dependence.\cite{K3} The susceptibility of CaCo$_{2}$As$_{2}$
displays a peak at 76 K. Our earlier paper had reported that
CaCo$_{2}$As$_{2}$ undergoes an antiferromagnetic transition with
the magnetic moments being aligned parallel to the \emph{c}
axis.\cite{Co-C} CaFe$_{2}$As$_{2}$ and CaCo$_{2}$As$_{2}$ will
enter into antiferromagnetic ordering states at low temperature.
However, the susceptibility of CaCo$_{2}$As$_{2}$ is about 30 times
larger than CaFe$_{2}$As$_{2}$. This distinct difference may be
ascribed to the different mechanisms of their antiferromagnetic
ordering. CaCo$_{2}$As$_{2}$ shows a characteristic itinerant
antiferromagnetism, and its in-plane ferromagnetism can be well
understood by a mean-field stoner instability. However, the magnetic
mechanism of CaFe$_{2}$As$_{2}$ is complex and is still under
debate. A simple itinerant picture can not well account for the
antiferromagnetism in CaFe$_{2}$As$_{2}$. CaNi$_{2}$As$_{2}$
exhibits a typical metal paramagnetic behavior and its
susceptibility shows little temperature dependence. The
susceptibility of CaCu$_{2}$As$_{2}$ has a negative sign and is
nearly no temperature dependence at the temperature range from 50 K
to 300 K. It seems that the transition discovered in resistivity
does not obviously affect the magnetic behavior of
CaCu$_{2}$As$_{2}$. At very low temperature, the susceptibilities of
CaNi$_{2}$As$_{2}$ and CaCu$_{2}$As$_{2}$ show upturns and deviate
from the temperature independence. These anomalies should not be the
intrinsic magnetic behavior of these samples and may originate from
the magnetic or paramagnetic impurities which may be introduced in
the process of sample growths.

\begin{figure}[t]
\scalebox{0.59} {\includegraphics [bb=330 30 6cm 16.5cm]{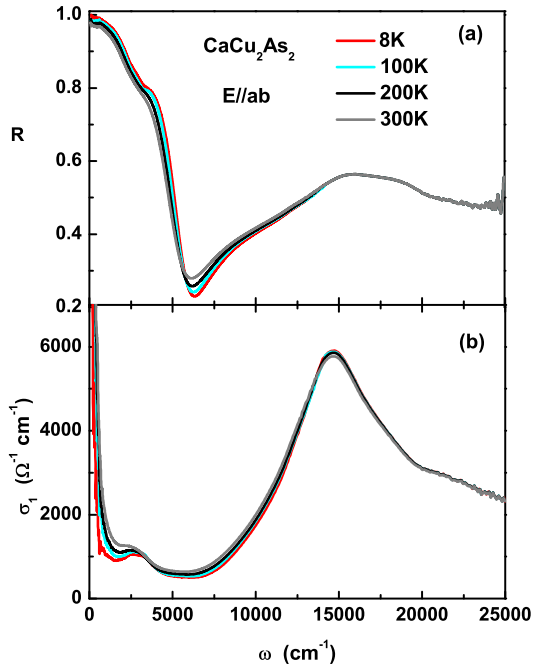}}
\caption{(Color online)(a) Temperature dependent optical reflectance
and (b) temperature dependent optical conductivity of
CaCu$_{2}$As$_{2}$.}
\end{figure}

The reflectance and the real part of optical conductivity of
CaCu$_{2}$As$_{2}$ are shown in Fig. 3(a) and 3(b), and a good
metallic behavior is exhibited both in frequency and temperature.
Different from many iron pnictides, a very sharp plasma edge at 5500
cm$^{-1}$ is observed in \emph{R($\omega$)}. Furthermore, a broad
dip, which is located at 2500 cm$^{-1}$, is displayed in the
room-temperature \emph{R($\omega$)} and becomes more pronounced as
temperature decreases. In the real part of optical conductivity, a
narrow Drude response is observed below 1500 cm$^{-1}$ and becomes
sharper with temperature cooling down. The small peak at 2500
cm$^{-1}$ relates to the broad dip observed in \emph{R($\omega$)}.
Optical measurements reveal that the optical conductivity of
CaCu$_{2}$As$_{2}$ does not develop any new energy gap across the
transition temperature. Combination with the hysteresis behavior of
resistivity, we assert that the anomaly discovered in resistivity of
CaCu$_{2}$As$_{2}$ can not be induced by any density-wave phase
transition. In CaFe$_{2}$(As$_{0.945}$P$_{0.055}$)$_{2}$ and
Ca$_{1-x}$\emph{Re}$_{x}$Fe$_{2}$As$_{2}$ (\emph{Re} = rare-earth
element), similar transport behavior has been observed by several
different groups and the inducement is regarded as the \emph{c} axis
abrupt collapse transition.\cite{Ca-P,Ca-Re} So we also regard that
the strange transport behavior discovered in CaCu$_{2}$As$_{2}$ may
originate from the simialar lattice abrupt collapse. The
low-temperature diffraction experiment is needed to confirm this
conjecture.

\begin{figure}[b]

\scalebox{0.55} {\includegraphics [bb=330 30 7cm
11.5cm]{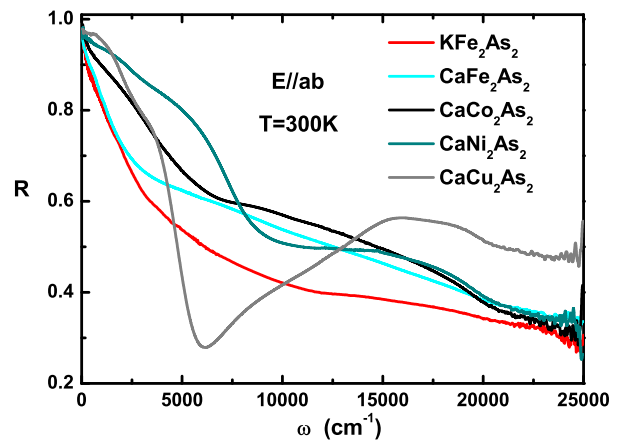}}

\caption{(Color online)  Optical reflectance of KFe$_{2}$As$_{2}$
and CaT$_{2}$As$_{2}$ (T = Fe, Co, Ni, Cu) at room temperature. }

\end{figure}

\begin{figure*}[t]

\includegraphics[clip,width=2.3in]{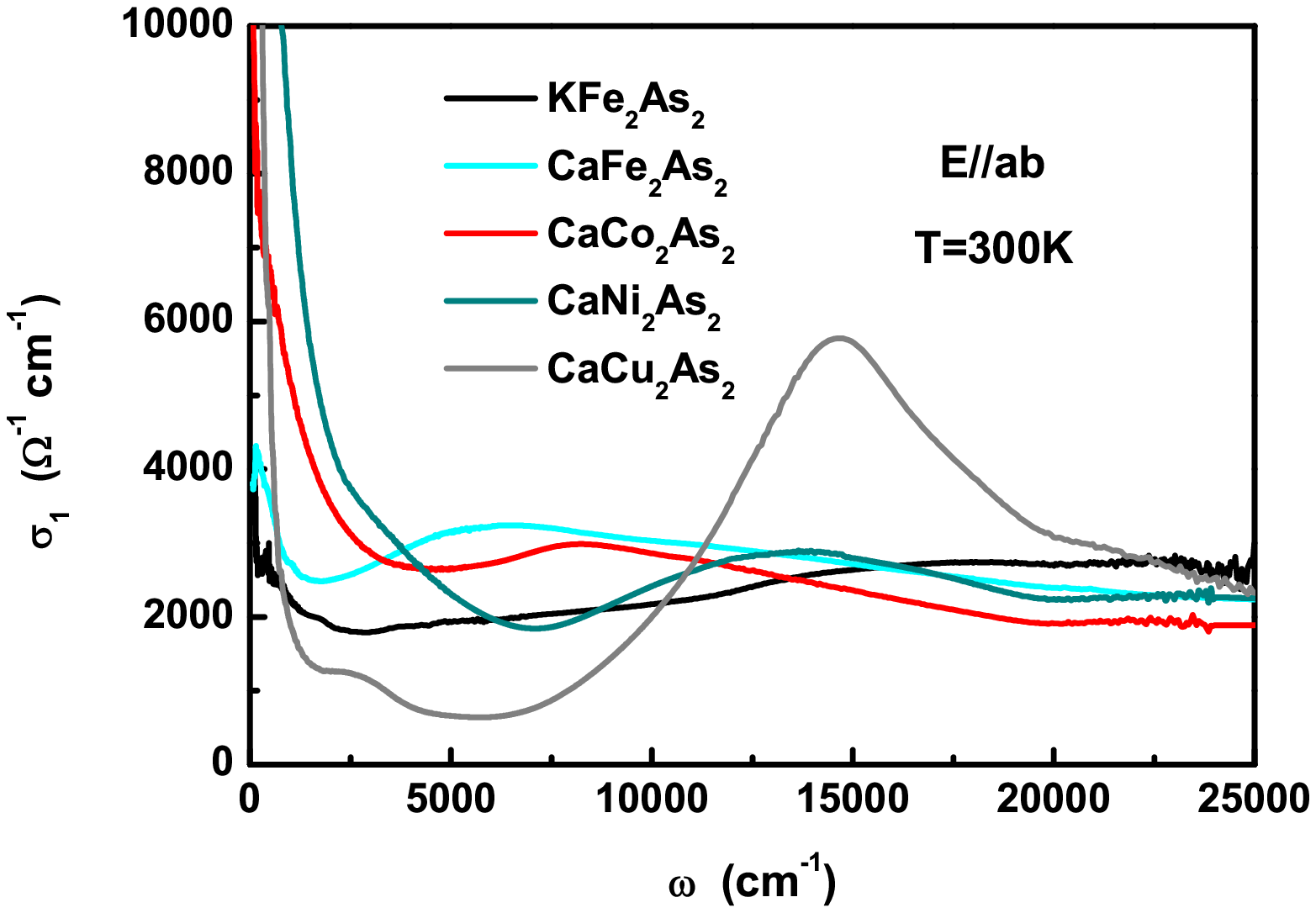}
\includegraphics[clip,width=2.3in]{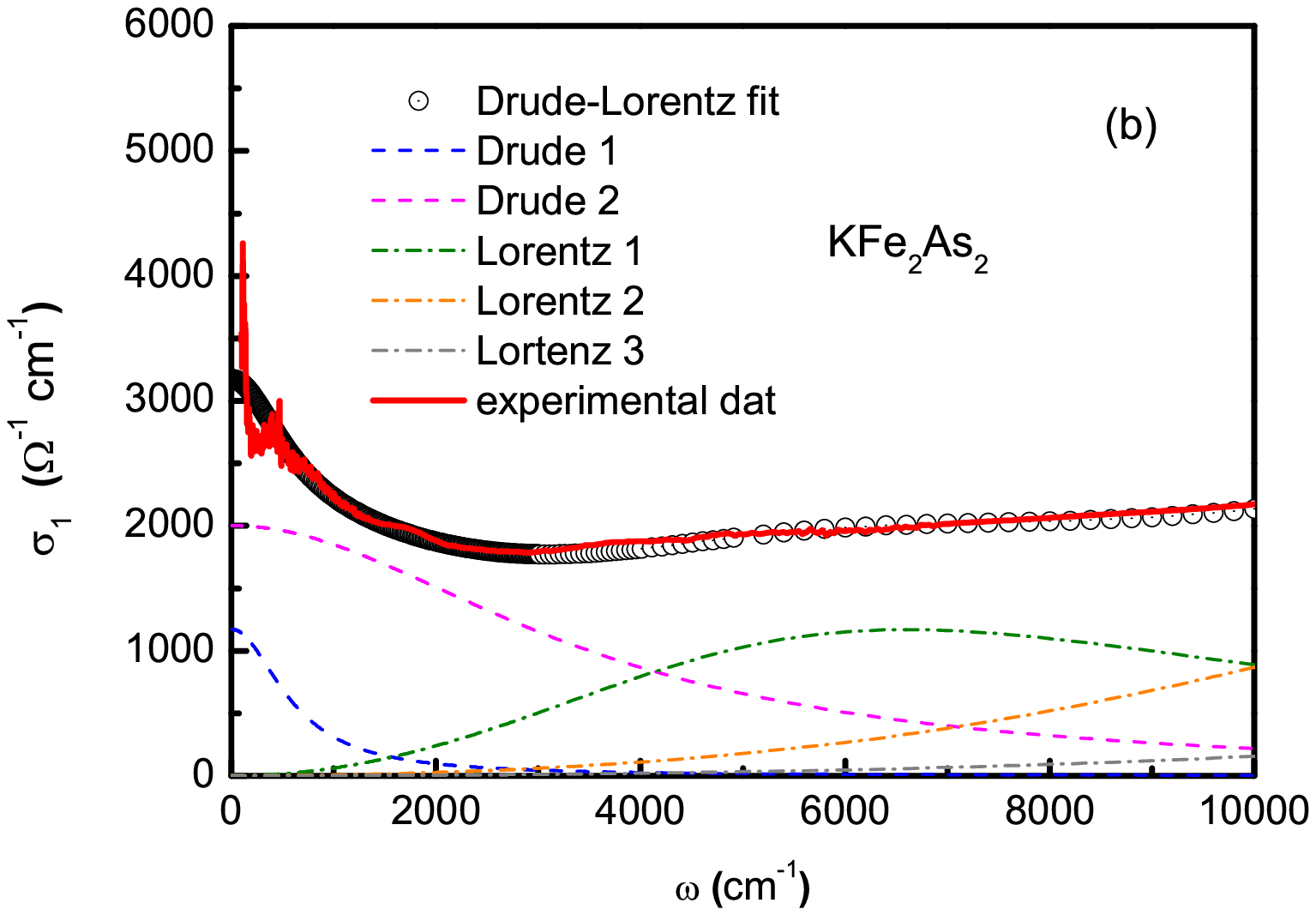}
\includegraphics[clip,width=2.3in]{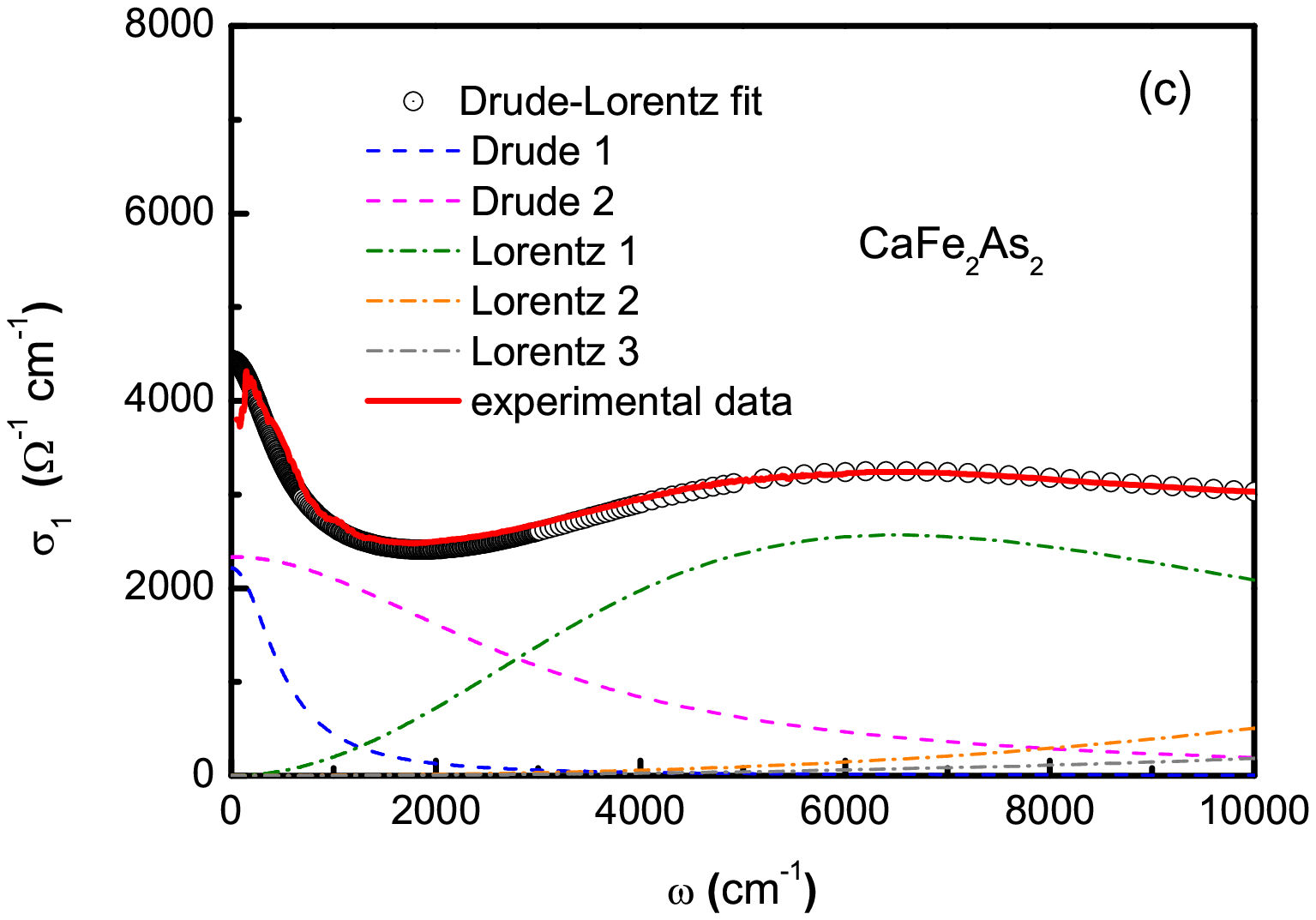}
\includegraphics[clip,width=2.3in]{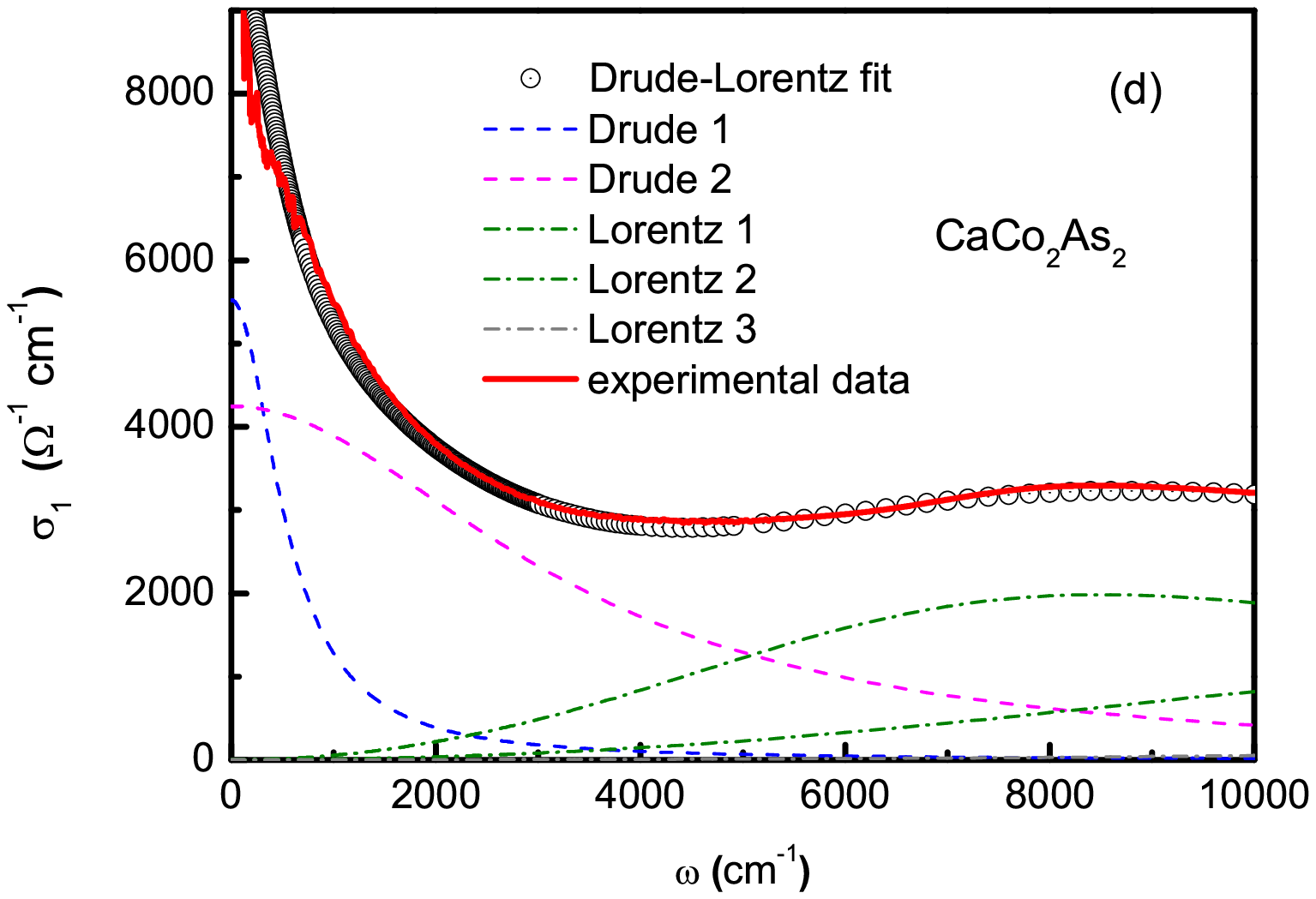}
\includegraphics[clip,width=2.3in]{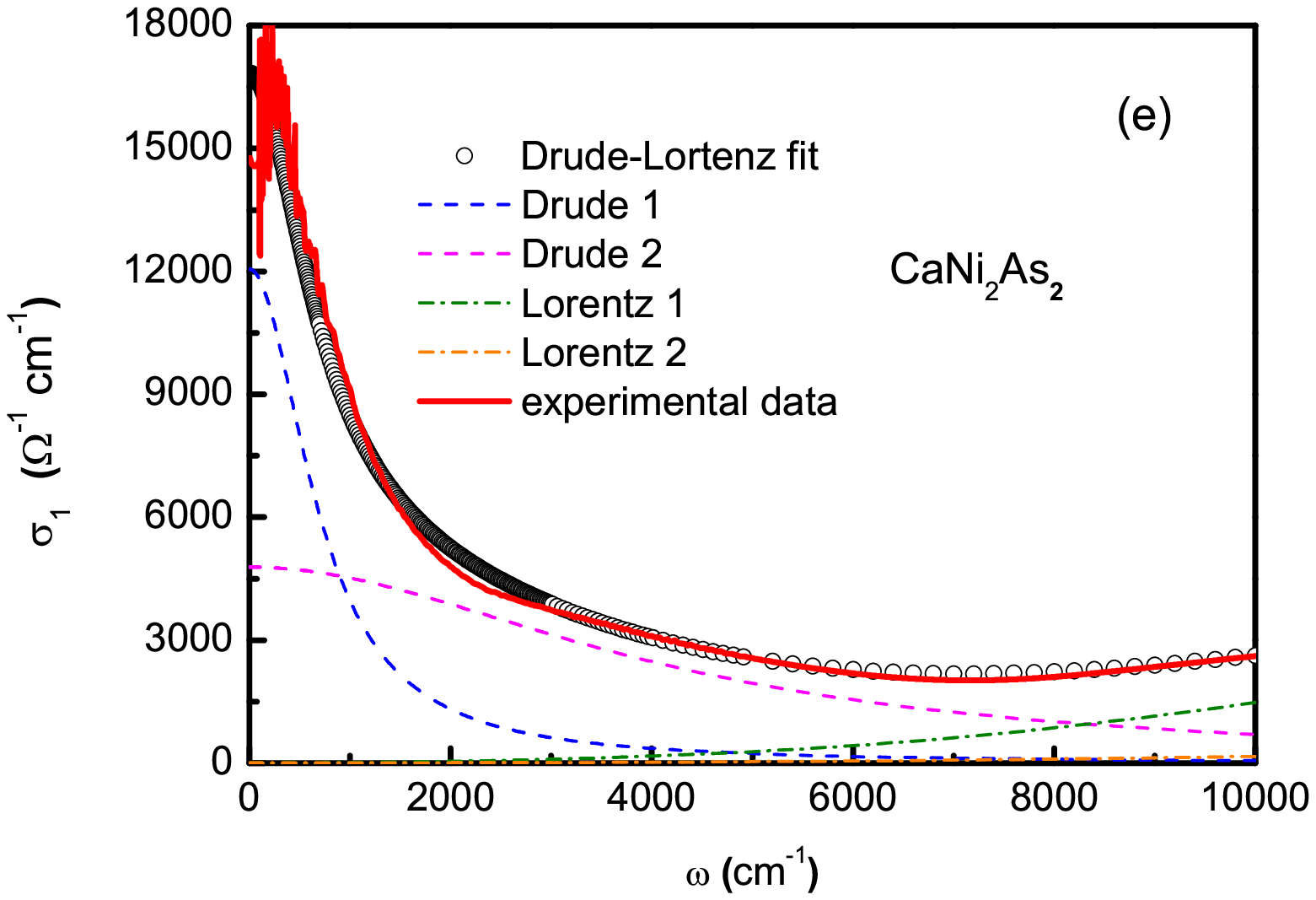}
\includegraphics[clip,width=2.3in]{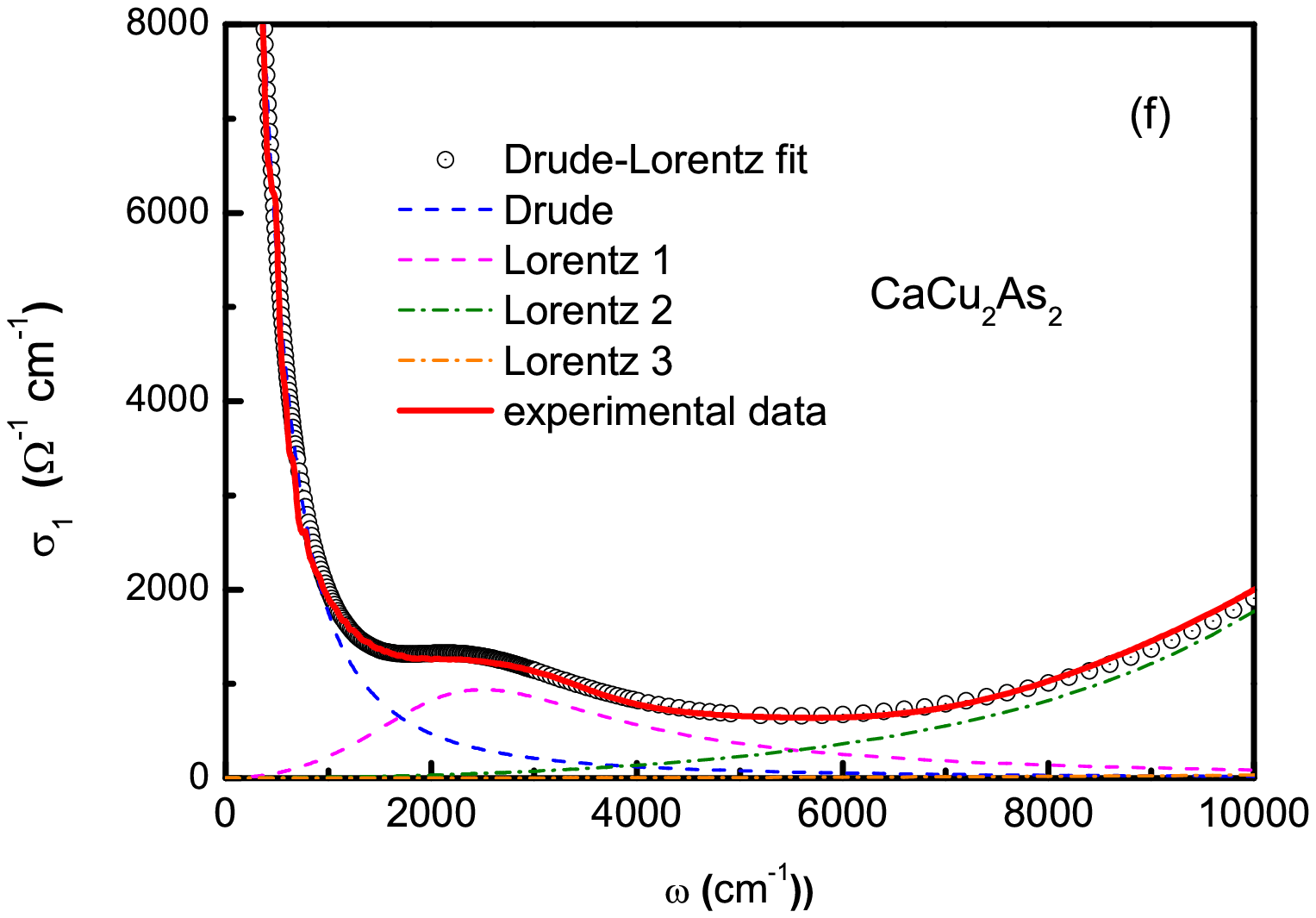}
\caption{(color online)  (a) Optical conductivity of
KFe$_{2}$As$_{2}$ and CaT$_{2}$As$_{2}$ (T = Fe, Co, Ni, Cu) at room
temperature. (b) Drude-Lorentz fit to the optical conductivity of
KFe$_{2}$As$_{2}$. (c) Drude-Lorentz fit to the optical conductivity
of CaFe$_{2}$As$_{2}$. (d) Drude-Lorentz fit to the optical
conductivity of CaCo$_{2}$As$_{2}$. (e) Drude-Lorentz fit to the
optical conductivity of CaNi$_{2}$As$_{2}$. (f) Drude-Lorentz fit to
the optical conductivity of CaCu$_{2}$As$_{2}$.}
\end{figure*}

Figure 4 shows the optical reflectance of KFe$_{2}$As$_{2}$ and
CaT$_{2}$As$_{2}$ (T = Fe, Co, Ni, Cu) at room temperature. The line
shapes of \emph{R($\omega$)} undergo drastic changes from one sample
to another, indicative of the notable changes of the band structures
with varying the nominal average number of \emph{3d }electrons on
per transition metal ion. Below 2500 cm$^{-1}$, \emph{R($\omega$)}
of KFe$_{2}$As$_{2}$ and CaT$_{2}$As$_{2}$ (T = Fe, Co, Ni)
approximately follow linear-$\omega$ dependence and plasma edges can
not be distinguished. However, CaCu$_{2}$As$_{2}$ shows a
pronouncedly different behavior from KFe$_{2}$As$_{2}$ and
CaT$_{2}$As$_{2}$ (T = Fe, Co, Ni). The reflectance of
CaCu$_{2}$As$_{2}$ exhibits a very sharp plasma edge at 5500
cm$^{-1}$. This sharp plasma edge is unexpected in a moderate or a
strong correlated system such as iron pnictides and the cuprates.
The real part of complex optical conductivity of these five samples
are shown in Fig. 5(a). Above 5000 cm$^{-1}$, the locations of
Lorentz peaks vary with different samples. Low-energy spectral
weight of $\sigma$$_{1}$($\omega$) gradually increases from
KFe$_{2}$As$_{2}$ to CaNi$_{2}$As$_{2}$. It is consistent with the
simple expectation that much more 3\emph{d} electrons will be
contributed to the density of states at Fermi level when the
transition metal ions have more nominal 3\emph{d} electrons. In the
conductivity spectra of CaCu$_{2}$As$_{2}$, a narrow Drude response
is observed. We use Drude-Lorentz model to decompose optical
conductivity:

\begin{equation}
\epsilon(\omega)=\epsilon_\infty-\sum_{s=1}^{M}{{\omega_{ps}^2}\over{\omega^2+i\omega/\tau_{Ds}}}+\sum_{j=1}^N{{S_j^2}\over{\omega_j^2-\omega^2-i\omega/\tau_j}}.
\label{chik}
\end{equation}

Here, $\epsilon_\infty$ is the dielectric constant at high energy,
the middle and last terms are the Drude and Lorentz components
respectively. We find that using two Drude terms, a narrow one and a
broad one, can fit the low-energy optical conductivity of
KFe$_{2}$As$_{2}$ and CaT$_{2}$As$_{2}$ (T = Fe, Co, Ni) very well.
But in CaCu$_{2}$As$_{2}$, the low-energy optical conductivity can
be well described by a narrow Drude term. Figure. 5(b) to 5(f) show
the fitting results of these five samples below 10000 cm$^{-1}$, and
the fitting parameters in the low-energy region are shown in the
Table I. From KFe$_{2}$As$_{2}$ to CaNi$_{2}$As$_{2}$, the spectral
weight of these two Drude terms simultaneously increase, and the
increase rate of the spectral weight of the narrow one is faster
than the broad Drude term. We plot the sample-dependent ratios of
narrow Drude spectral weight in the total Drude spectral weight in
Fig. 6. From KFe$_{2}$As$_{2}$ to CaNi$_{2}$As$_{2}$, the ratio of
narrow Drude spectral weight gradually increases, and its value
keeps at the range of 0.1 to 0.3. However, in CaCu$_{2}$As$_{2}$,
the ratio of narrow Drude spectral weight reaches up to 1, which is
much higher than other four samples. These obviously different
results supply a clue for us to understand the universal two Drude
terms decomposition of low-energy optical conductivity and the
origin of the broad Drude terms in iron pnictides. However, although
the spectral weight changes a lot, the scattering rates of these two
Drude terms basically show little sample dependence. A notable
feature of CaCu$_{2}$As$_{2}$ is that the scattering rate of its
Drude term is much smaller than the scattering rates of the narrow
Drude terms of other four samples.

\begin{figure}[b]
\scalebox{0.45} {\includegraphics [bb=600 30 -1.5cm 13cm]{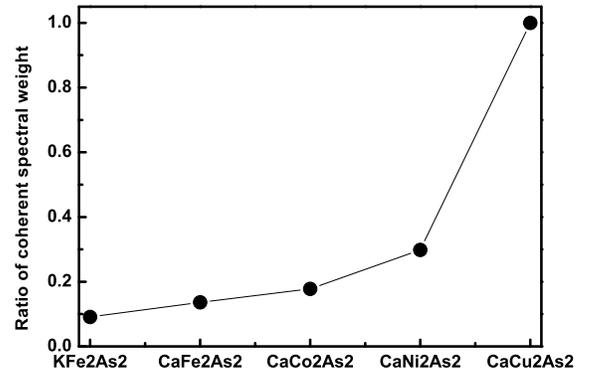}}
\caption{ Sample dependent ratio of coherent Drude spectral weight
in the total Drude spectral weight }
\end{figure}

\begin{table*}[t]
\begin{center}
\newsavebox{\tablebox}
\begin{lrbox}{\tablebox}
\begin{tabular}{*{15}{m{16mm}}}
\hline\\[0.5pt]
{}&$\omega_{p1}$&$\gamma_{D1}$&$\omega_{p2}$&$\gamma_{D2}$ &$\omega_{c}$&$\omega_{p}$&$\sqrt{\omega_{p1}^{2}+\omega_{p2}^{2}}$\\
\hline\\[0.5pt]
KFe$_{2}$As$_{2}$&6500&600&20500&3500&3000&15800&21500\\[4pt]
CaFe$_{2}$As$_{2}$&8150&500&20500&3000&2000&15000&22000\\[4pt]
CaCo$_{2}$As$_{2}$&13500&550&29000&3300&4000&26000&32000\\[4pt]
CaNi$_{2}$As$_{2}$&22500&680&34500&4150&6000&35000&41200\\[4pt]
CaCu$_{2}$As$_{2}$&19000&320&--&--&1500&18000&19000\\[4pt]
\hline\\[5pt]
\end{tabular}
\end{lrbox}
\caption{Fitting parameters of Drude terms in KFe$_{2}$As$_{2}$ and
CaT$_{2}$As$_{2}$ (T = Fe, Co, Ni, Cu) at 300 K. $\omega$$_{p1}$ and
$\omega$$_{p2}$ are the plasma frequency of the narrow Drude term
and the broad Drude term. $\gamma$$_{D1}$ and $\gamma$$_{D2}$ are
the scattering rate of the narrow Drude term and the broad Drude
term. $\omega_{c}$ is the cutoff frequency. $\omega_{p}$ is the
plasma frequency estimated via an integral of the real part of the
optical conductivity up to a cutoff frequency
$\omega$$_{c}$.$\sqrt{\omega_{p1}^{2}+\omega_{p2}^{2}}$ is the
plasma prequency estimated via Drude-Lorentz fit. The unit of these
quantities is cm $^{-1}$.} \scalebox{1}{\usebox{\tablebox}}
\end{center}
\end{table*}

It should be emphasized that the two-Drude-term decomposition is not
the only way to reproduce the low-energy optical conductivity of
iron pnictides. Furthermore, the broad Drude term is lack of
definitive physical meaning. It extends into near-infrared region
and obviously contains spectral weight of interband transition,
which will result in a systemic overestimation of the plasma
frequencies of iron pnictides. However, some other methods, which
have been used to describe the low-energy optical conductivity of
iron pnictides, also have problems to estimate the plasma frequency.
For example, if a narrow Drude term and a Lorentz term located at
low frequency are used to describe the low-energy optical
conductivity,\cite{BaCo1} the plasma frequences of iron pnictides
are always underestimated. Furthermore, the physical meaning of this
Lorentz term is also unclear. It is hard to believe that the
localization effects induced by impurties and the interband
transition will contribute so many spectral weight to the
far-infrared region. All these facts indicate that no matter which
method is used to analyze the experimental data, the accurate border
between intraband transition and interband transition is difficult
to be circumscribed by using Drude-Lorentz model. To reveal the fact
that Drude-Lorentz fit can not give a reasonable estimation of the
plasma frequency of these samples, we also present the plasma
frequency in Table I estimated via an integral of the real part of
the optical conductivity up to a cutoff frequency $\omega$$_{c}$:
\begin{equation}
\omega_{p}^{2}=Z_{0}/\pi^{2}\int_{0}^{\omega_{c}}\sigma_{1}(\omega)\emph{d}\omega.
\label{chik}
\end{equation}

where Z$_{0}$ is the vacuum impedance, and the cutoff frequency
$\omega$$_{c}$ is determined by the first minimum of the real part
of optical conductivity. This method is always used to estimate the
spectral weight of Drude responses and is regarded as a reasonable
way to obtain plasma frequency of iron pnictides.\cite{LaP} It can
be seen that, except CaCu$_{2}$As$_{2}$, the plasma frequency of the
other four samples obtained by integration are clearly smaller than
the value given by Drude-Lorentz fit. For instance, the plasma
frequency of KFe$_{2}$As$_{2}$ obtained through integration is 15800
cm$^{-1}$, but the value given by fitting is about 21500 cm$^{-1}$.
The density function calculations reveal the plasma frequency of
KFe$_{2}$As$_{2}$ is about 21000 cm$^{-1}$.\cite{LDA3,LDA} Taking
the mass renormalization estimated via dynamical mean field theory
calculation (DMFT) into account,\cite{DMFT3} the actual plasma
frequency of KFe$_{2}$As$_{2}$ should be much smaller than 21000
cm$^{-1}$, and 13000 cm$^{-1}$ or below may be an accepted value.
The fitting value of plasma frequency is obviously larger than 13000
cm$^{-1}$ and clearly indicates the decomposition of low-energy
optical conductivity into two Drude terms indeed bring about the
overestimation about the plasma frequency. At the same time, the
plasma frequency estimated via integration is also sightly larger
than 13000 cm$^{-1}$. It is well known that there exist several
factors which will affect the accuracy of estimation about the
plasma frequency via integration. The first factor is the accuracy
about the measurements and the high-energy extrapolation to obtain
optical conductivity, and the second factor is the accuracy of the
estimation about the cutoff frequency $\omega_{c}$. The
$\sigma_{1}(\omega)$ of KFe$_{2}$As$_{2}$ is too flat between 2000
cm$^{-1}$ and 4000 cm$^{-1}$, bringing about the difficulty to
choose a suitable cutoff frequency which will balance between the
onset of the spectral weight of interband transition and the tail of
the Drude component. Not only KFe$_{2}$As$_{2}$, CaCo$_{2}$As$_{2}$
and CaNi$_{2}$As$_{2}$ also have this problem. Taking these factors
into account, 15800 cm$^{-1}$ may be an accept value from the
optical experimental side. Here, we emphasize that our motivation to
use Drude-Lorentz model to decompose the real part of optical
conductivity is not to get a reasonable plasma frequency of these
samples.We synthesize five selective samples and perform optical
studies on them. By using a similar procedure to decompose the
low-energy optical conductivity, we try to find out why the plasma
frequency of iron pnictides can not be estimated correctly through
Drude-Lorentz model and why the incoherent Drude term is always
present in the optical conductivity of iron pnictides.

Earlier optical investigation reveals that decomposition of the
low-energy optical conductivity into two Drude terms, a narrow one
and a broad one, is a universal behavior in iron
pnictides.\cite{BaCo1,TD,BaCo2} The broad Drude term has a long
high-energy tail, indicative of its incoherent feature. The nesting
condition of Fermi surfaces in iron pnictides, which may induce
strong scattering between the quasi-particles on hole and electron
Fermi surfaces, is a potential candidate for the interpretation to
this incoherent Drude term.\cite{BaCo2} According to our data, the
transitional metal ions in CaCo$_{2}$As$_{2}$ and CaNi$_{2}$As$_{2}$
have one more and two more 3\emph{d} electrons than Fe ions in
CaFe$_{2}$As$_{2}$. The Fermi surfaces of CaCo$_{2}$As$_{2}$ and
CaNi$_{2}$As$_{2}$ evolve into more complex shapes and will not
fulfill the nesting condition for an SDW instability. However, the
incoherent Drude term also can be subtracted from
$\sigma$$_{1}$($\omega$) of CaCo$_{2}$As$_{2}$ and
CaNi$_{2}$As$_{2}$, and the incoherent Drude term has large spectral
weight in the total Drude spectral weight. These results rule out
the proposal that the incoherent term originates from the effects of
some Fermi-surface sheets or their segments fulfilling the nesting
condition for SDW instability. Furthermore, drastic changes of Fermi
surfaces from KFe$_{2}$As$_{2}$ to CaNi$_{2}$As$_{2}$ make it is
very difficult to assert that the two different Drude terms in
KFe$_{2}$As$_{2}$ and CaT$_{2}$As$_{2}$ (T=Fe, Co, Ni) originate
from two different types of quasi-particles on the hole and electron
Fermi surfaces. Band structure calculation reveals that parent
compounds of iron-based superconductors have two electron Fermi
surfaces centered at the folded Brillouin zone (FBZ) corner and
three hole Fermi surfaces centered at FBZ center in their
paramagnetic states, and the quasi-particles on the hole and
electron Fermi surfaces may have different Fermi
velocities.\cite{LDA1,LDA2} Hole (Electron) doping will expand
(shrink) the hole Fermi surfaces and shrink (expand) the electron
Fermi surfaces in iron pnictides. For instance, the heavy hole
doping into BaFe$_{2}$As$_{2}$ drastically changes its band
structure. Angle-resolved photoemission spectroscopy (ARPES)
measurements reveal that the electron Fermi surfaces which will
exist in the parent and the optimally doped compounds are completely
absent in KFe$_{2}$As$_{2}$.\cite{KFA} Furthermore, the heavy
electron doping, such as in CaCo$_{2}$As$_{2}$ and
CaNi$_{2}$As$_{2}$, aslo drastically changes band structures. Band
structure calculation reveals that the electron Fermi surfaces are
dominant in these compounds.\cite{Co5,Ni} However, although the two
types of Fermi surfaces vary severely with different samples, the
ratio of coherent Drude spectral weight dose not show notable
changes. These results mean that the two different types of Drude
terms can not originate from different optical response of hole and
electron Fermi surfaces.

\begin{figure}[t]
\scalebox{0.45} {\includegraphics [bb=600 120 -2cm 16.5cm]{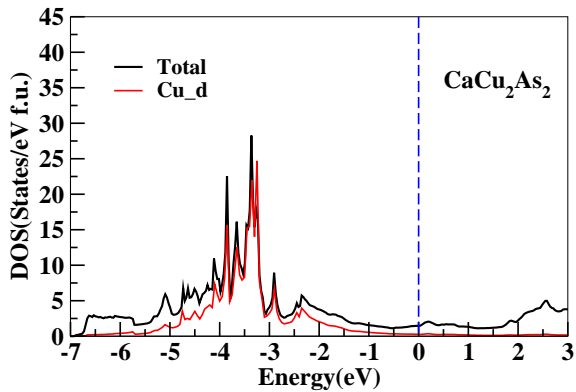}}
\caption{(Color online) Electronic density of states and Cu
3\emph{d} projection for CaCu$_{2}$As$_{2}$. }
\end{figure}

Figure 7 shows the calculated density of states of
CaCu$_{2}$As$_{2}$. Different from KFe$_{2}$As$_{2}$ and
CaT$_{2}$As$_{2}$ (T = Fe, Co, Ni), the Cu 3\emph{d} orbitals are
mainly located below the Fermi energy and basically do not
contribute to the density of states at the Fermi level. The
4\emph{s} and 4\emph{p} orbitals of As and Cu are mainly responsible
to the transport properties and low-energy optical response. It is
well known that the 4\emph{s} and 4\emph{p} orbital wave functions
are more spatially extended than 3\emph{d} orbital wave functions,
and show much weaker electronic correlation effects than 3\emph{d}
electrons. The sharp plasmas edge and the coherent Drude responses
of \emph{R($\omega$)} and $\sigma$$_{1}$($\omega$) in
CaCu$_{2}$As$_{2}$ may have relations with this weak electronic
correlation. For the Hubbard and \emph{t}-\emph{J} model in two
dimension, calculation with dynamical mean field theory reveals that
the electronic correlation effects usually push the Drude response
into the incoherent side.\cite{DMFT} Especially in some strong
electronic correlation materials such as the cuprates and Ni, Ti, V,
Mn-based ABO$_{3}$-structure compounds, the low-energy
$\sigma$$_{1}$($\omega$) always can not be fitted well by using only
one narrow Drude term, and the incoherent background of
$\sigma$$_{1}$($\omega$) at low frequency is always described by a
bound excitation with a Lorentz term centered at low
frequency.\cite{S1,S2,S3} Iron pnictides are regraded as moderate
electronic correlation systems and the Hubbard \emph{U} is estimated
to be 2$\sim$4 eV.\cite{SW} So it is very reasonable to expect that
the moderate electronic correlation will generate visible effects on
the low-energy optical response. CaCo$_{2}$As$_{2}$ and
CaNi$_{2}$As$_{2}$ are also the 3\emph{d}-electron correlated
systems and have similar crystal structures with CaFe$_{2}$As$_{2}$.
Their low-energy optical response share similar features with
KFe$_{2}$As$_{2}$ and CaFe$_{2}$As$_{2}$. According to these facts,
we conjecture that the common incoherent Drude optical response in
KFe$_{2}$As$_{2}$ and CaT$_{2}$As$_{2}$ (T = Fe, Co, Ni) may
originate from the moderated electronic correlated effects and will
not depend on the details of band structures, causing the difficulty
to circumscribe the accurate border between intraband transition and
interband transition.

\begin{figure}[t]
\scalebox{0.45} {\includegraphics [bb=600 25 -1cm 13cm]{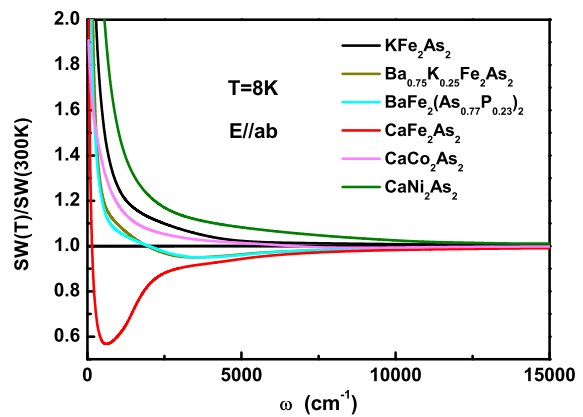}}
\caption{(Color online) Ratio of the integrated spectral weight as a
function of $\omega$ at 8 K in KFe$_{2}$As$_{2}$,
Ba$_{0.75}$K$_{0.25}$Fe$_{2}$As$_{2}$,
BaFe$_{2}$(As$_{0.77}$P$_{0.23}$)$_{2}$ and CaT$_{2}$As$_{2}$ (T =
Fe, Co, Ni). The low temperature integrated spectral weight is
normalized to the room-temperature data.}
\end{figure}

In Fig. 8, we plot integrated spectral weight of these samples at 8
K, and the low temperature integrated spectral weight has been
normalized to the integrated spectral weight at room temperature.
For the convenience of comparison, we also present the
low-temperature integrated spectral weight of
Ba$_{0.75}$K$_{0.25}$Fe$_{2}$As$_{2}$ and
BaFe$_{2}$(As$_{0.77}$P$_{0.23}$)$_{2}$ which we have never
published before. In the integrated spectral weight of
Ba$_{0.75}$K$_{0.25}$Fe$_{2}$As$_{2}$,
BaFe$_{2}$(As$_{0.77}$P$_{0.23}$)$_{2}$ and CaFe$_{2}$As$_{2}$,
there is a clear transfer of spectral weight (SW) from low to high
energy with decreasing temperature, which is similar to the earlier
reports about BaFe$_{2}$As$_{2}$ and
Ba(Fe$_{1-x}$Co$_{x}$)$_{2}$As$_{2}$.\cite{SW2,SW} It is believed
that this unconventional spectral weight transfer has relations with
the Hund's coupling energy \emph{J$_{H}$}, which is estimated to be
0.6 $\thicksim$ 0.9 eV.\cite{SW2,SW} However, in the integrated
spectral weight of KFe$_{2}$As$_{2}$, CaCo$_{2}$As$_{2}$ and
CaNi$_{2}$As$_{2}$, this unconventional spectral weight transfer can
not be observed. The lack of the unconventional spectral weight
transfer in CaCo$_{2}$As$_{2}$ and CaNi$_{2}$As$_{2}$ indicates that
the \emph{J$_{H}$}-related spectral weight transfer only displays
visibly in the materials which show exotic magnetism and
superconductivity in iron pnictides, and \emph{J$_{H}$} may play
important roles in superconductivity and magnetism in iron
pnictides.

\section{\label{sec:level2}SUMMARY}

In summary, a systematic investigation of resistivity,
susceptibility and optical spectroscopy are performed on
KFe$_{2}$As$_{2}$, CaFe$_{2}$As$_{2}$, CaCo$_{2}$As$_{2}$,
CaNi$_{2}$As$_{2}$, and CaCu$_{2}$As$_{2}$. we find
CaCu$_{2}$As$_{2}$ undergoes a transition at 50 K, which is similar
to the lattice abrupt collapse transition discovered in
CaFe$_{2}$(As$_{1-x}$P$_{x}$)$_{2}$ and
Ca$_{1-x}$\emph{Re}$_{x}$Fe$_{2}$As$_{2}$ (\emph{Re} = rare-earth
element). However, optical measurements reveal that the optical
response of CaCu$_{2}$As$_{2}$ is not very sensitive to the
transition at 50 K. Resistivity and susceptibility studies reveal
that although these samples have very different magnetic properties,
they share some common features in transport properties. Using
Drude-Lorentz model to analyze $\sigma$$_{1}$($\omega$) of
KFe$_{2}$As$_{2}$ and CaT$_{2}$As$_{2}$ (T = Fe, Co, Ni, Cu), we
find that using two Drude terms, a coherent one and an incoherent
one, can fit the low-energy $\sigma$$_{1}$($\omega$) of
KFe$_{2}$As$_{2}$ and CaT$_{2}$As$_{2}$ (T = Fe, Co, Ni) very well.
However, in CaCu$_{2}$As$_{2}$, one coherent Drude term can account
for the low-energy $\sigma$$_{1}$($\omega$) well. Lack of the
incoherent Drude term in CaCu$_{2}$As$_{2}$ may be attributed to the
weaker electronic correlation compared to KFe$_{2}$As$_{2}$ and
CaT$_{2}$As$_{2}$ (T = Fe, Co, Ni). We also perform spectral weight
analysis on these samples. We find that the unconventional spectral
weight transfer related to Hund's coupling energy \emph{J$_{H}$} is
only observed in iron pnictides, indicative of that \emph{J$_{H}$}
may play an important role in the mechanism of magnetism and
superconductivity in pnictides.

\begin{center}
\small{\textbf{ACKNOWLEDGMENTS}}
\end{center}
 This work was supported by the National Science Foundation of China (10834013,
11074291) and the 973 project of the Ministry of Science and
Technology of China (2012CB821403)

\end{document}